\documentclass[12pt,a4paper]{article}
\pdfoutput=1
\usepackage{cite}
\usepackage{amsmath, amsthm, amssymb, graphicx,slashed}
\usepackage{jheppub}
\usepackage{etoolbox}
\makeatletter
\patchcmd{\maketitle}{\@fpheader}{}{}{}
\makeatother

\numberwithin{equation}{section}

\newcommand{\be}{\begin{equation}}
\newcommand{\ee}{\end{equation}}
\newcommand{\bea}{\begin{eqnarray}}
\newcommand{\eea}{\end{eqnarray}}

\newcommand{\nn}{\nonumber}

\renewcommand{\bar}{\overline}

\def\bal{\begin{array}{l}}
\def\ba#1{\begin{array}{#1}} 
\def\ea{\end{array}}
\def\beas{\begin{eqnarray*}}
\def\eeas{\end{eqnarray*}}

\def\eq#1{(\ref{#1})}

\def\nn{\\\nonumber}

\def\bit{\begin{item}}
\def\eit{\end{item}}
\def\benu{\begin{enumerate}}
\def\eenu{\end{enumerate}}

\makeatletter
\@addtoreset{equation}{section}
\makeatother

\newcommand{\comment}[1]{}

\def\p{\partial}

\def\ba{\begin{align}}
\def\ea{\end{align}}
\def\beq{\begin{eqnarray}}
\def\eeq{\end{eqnarray}}

\title{\boldmath Hot Attractors}

\author{Kevin Goldstein,}
\author{Vishnu Jejjala,}
\author{Suresh Nampuri}

\affiliation{Mandelstam Institute for Theoretical Physics, School of Physics, and National Institute for Theoretical Physics, University of the Witwatersrand, Johannesburg, WITS 2050, South Africa}

\emailAdd{kevin.goldstein@wits.ac.za}
\emailAdd{vishnu@neo.phys.wits.ac.za}
\emailAdd{suresh.nampuri@wits.ac.za}

\abstract{
  The product of the areas of the event horizon and the Cauchy horizon
  of a non-extremal black hole equals the square of the area of the
  horizon of the black hole obtained from taking the smooth extremal limit.
  We establish this result for a large class of black holes using the
  second order equations of motion, black hole thermodynamics, and the attractor mechanism for
  extremal black holes.
  This happens even though the area of each horizon generically depends on the moduli, which are asymptotic values of scalar fields.
  The conformal field theory dual to the BTZ black hole facilitates a microscopic interpretation of the result.
  In addition, we demonstrate that certain quantities which vanish in the extremal case are zero when integrated over the region between the two horizons.
  We corroborate these conclusions through an analysis of known solutions.
}

\preprint{{\footnotesize WITS-CTP-142}}
\begin{document}

\maketitle


\section{\label{sec:intro}Introduction}

Black holes obey the familiar laws of thermodynamics~\cite{bh1,haw}.
In particular they have an entropy, which scales like one quarter of
the area of the event horizon.  Four decades after this realization,
the underlying theory of statistical physics that explains the
microscopic origin of entropy remains elusive.  While significant
progress has been made in cataloging microstates of maximally
supersymmetric, extremal black holes within string
theory~\cite{sv1,sv2,sv3,sv4}, a similar understanding of black holes
with a lower amount of symmetry is substantially less developed.
Identifying the microstates relies crucially on
holography~\cite{ts1,ts2} whose best studied avatar is the AdS/CFT
correspondence~\cite{adscft1,adscft2,adscft3}.  The presence of the
AdS$_3$ factor in the near-horizon limit of extremal systems at least
enables the application of the Cardy formula~\cite{cardy1,cardy2,cardy3} in the dual
CFT$_2$ to enumerate the microstates that
account for the entropy.  Surprisingly, this formula correctly
reproduces the entropy even when the temperature in the CFT is of
order one~\cite{str,kerrcft,jn}.

Charged non-extremal black holes have a least two horizons.
They are not protected by supersymmetry.
The first law relates the differential change in entropy of these black holes to the differential changes in ADM quantities such as mass, charge, and angular momentum.
The surface gravity at the outer horizon determines the temperature at which the non-extremal black hole radiates Hawking particles to infinity.
In an asymptotically flat space, this supplies a mechanism for black hole evaporation via thermal emission.
In suitable coordinates, the temperature is proportional to the distance between the inner and outer horizons.

The application of thermodynamical principles to the inner horizon
remains a work in progress~\cite{curir1,curir2,Cvetic:1997uw,hep-th/9708090}.  Assigning a
statistical meaning to the inner horizon is difficult for many
reasons, not least because the inner horizon is classically unstable
~\cite{penrose68,inst1,poisson90,inst2,Marolf:2010nd,inst3}.  A hint at the importance of the
inner horizon for improving our conception of black hole statistical
mechanics is nevertheless conveyed by the observation that the product
of the areas of the inner and outer Killing horizons is a function
only of the quantized charges and is independent of the mass of the
black hole~\cite{finn}.  Moreover, the geometric mean of the areas of
the inner and outer horizon yields the area of the black hole obtained
from taking the smooth extremal limit~\cite{cgp}:
\be
\sqrt{A_+\, A_-}
= A_\mathrm{ext} ~.
\label{eq:areas}
\ee
This is an empirical statement for which a geometric proof is so far lacking.

To start, let us consider black holes in three dimensions.
Using the analysis of Brown and Henneaux that the asymptotic symmetry group of AdS$_3$ is generated by two copies of the Virasoro algebra, the level matching condition in the dual theory ensures that
\be
\frac{A_+\, A_-}{(8\pi G_3)^2} = n_R - n_L ~,
\ee
where the right hand side expresses the difference in the number of right-moving and left-moving excitations in the CFT$_2$~\cite{bh2,str}.near-horizon
This is, of course, an integer independent of the mass.
The central charges on the left and the right are
\be
c_L = c_R = \frac{3L}{2G_3} ~,
\ee
where $L$ is the radius of anti-de Sitter space, and the gravity description applies in the limit where these charges are large~\cite{bh2}.

While it remains unclear what quantum mechanical degrees of freedom the area of the inner horizon counts, a first law of thermodynamics applies to this surface~\cite{cretc1,cretc2,cretc3,cretc4,cretc5,Chen:2012mh,Chen:2012yd,Chen:2013rb}.
For comparison, we write
\be
dM = T_+ dS_+ - \Omega_+ dJ \quad (\mathrm{outer}) ~, \qquad \qquad
dM = T_- dS_- - \Omega_- dJ \quad (\mathrm{inner}) ~,
\ee
where we use~\cite{ms} to cast the thermodynamic quantities as
\be
S_\pm = \frac{A_\pm}{4G_3} = \frac{\pi r_\pm}{2G_3} ~, \quad
T_\pm = \pm \frac{r_+^2-r_-^2}{2\pi r_\pm L^2} ~, \quad
J = \frac{r_+ r_-}{4G_3 L} ~, \quad
\Omega_\pm = \frac{r_\mp}{r_\pm L} ~.
\ee
In CFT language, we may equally express
\be
T_{R,L} = \frac{r_+\pm r_-}{2\pi L^2} ~, \quad
T_\pm^{-1} = \frac12 \left( T_R^{-1} \pm T_L^{-1} \right) ~, \quad
S_\pm = \frac{\pi^2 L}{3} (c_R T_R \pm c_L T_L) ~.
\ee
To preserve the form of the first law at the two horizons, in our notation, we take the temperature of the inner horizon to be negative.
In gravity, this is a choice that we have the freedom to make~\cite{barn}.
Crucially, the temperatures of the left movers and the right movers in the CFT are both positive.
These relations apply to the BTZ black hole~\cite{btz1,btz2} and have been discussed in the literature~\cite{cretc1,cretc2,cretc3,cretc4,cretc5,Chen:2012mh,Chen:2012yd,Chen:2013rb}.
To briefly summarize a few of the other known results, let us note that~\eq{eq:areas} encapsulates a statement about the thermodynamic entropy:
\be
S_+\, S_- = S_\mathrm{ext}^2 ~.
\label{eq:entropies}
\ee
This equality, which is true for the BTZ solution, also holds in $d=4,5$ for black hole geometries in Einstein--Maxwell theory that have flat, de Sitter, or anti-de Sitter asymptopia.
It also applies to certain examples in higher derivative gravity where the Wald entropy is proportional to the area.
The product of the entropies $S_+\, S_- \ne S_\mathrm{ext}^2$ when the Wald entropy is not proportional to area.
In such examples, the Smarr relations fail.
(In gravitational thermodynamics, the Smarr relations are statements of the Euler equation and the Gibbs--Duhem relation.)
Nevertheless, we have
\begin{equation}
  \label{eq:geom_area}
 A_+\, A_- = A_\mathrm{ext}^2
\end{equation}
 in these cases as well.

The black hole attractor mechanism, in which the extremal horizon
area, $A_\mathrm{ext}$ (or more generally the Wald entropy), is
independent of the asymptotic moduli, is a well known phenomenon~\cite{attr1,attr2,attr3,attr4,attr5,hep-th/0506177,hep-th/0507096,hep-th/0606244,hep-th/0611143,0708.1270}.
The mechanism ensures that moduli are drawn to fixed values on the
horizon independent of their asymptotic starting points (unless the
moduli correspond to flat directions which do not effect the entropy).
However, the attractor mechanism fails once the temperature is non-zero ---
generically $A_\pm$ are moduli dependent. We find, from rather simple
considerations, that for a large class of black holes
(\ref{eq:geom_area}) holds.  This means that the product of areas is
independent of the moduli, which suggests a non-local generalization of
the attractor mechanism to the non-extremal case involving both
horizons. In the extremal case the values of the moduli are determined
by attractor equations evaluated on the horizon (or equivalently in
the near-horizon geometry for the entropy function
formalism~\cite{hep-th/0506177}). We find a non-extremal generalization
where the attractor equations still hold once we average them over the
region between the horizons.

The organization of the paper is as follows.  In Section~\ref{sec:eom}
we review some relevant black hole mechanics, thermodynamics and the
attractor equations. In Section~\ref{sec:hot-attr-equat} we present
generalized non-extremal attractor equations and a proof
of~(\ref{eq:entropies}). In Section~\ref{sec:mod} we discuss features
of the moduli flow between the horizons for certain known solutions.
In Section~\ref{sec:cft} we discuss a CFT interpretation of a our
results. In Section~\ref{sec:disc} we discuss our conclusions and
possible future directions, and finally, in the Appendices we present
some technical details and additional plots.

\section{Equations of motion}
\label{sec:eom}
We are interested in non-extremal black holes whose extremal limit
displays attractor behavior.  In this section, we review some results
from \cite{hep-th/0507096,hep-th/0512138} and mention some minor generalizations to the non-extremal case.
We consider
four dimensional gravity coupled to $U(1)$ gauge fields and moduli,
\begin{equation}
  S=\frac{1}{\kappa^{2}}\int d^{4}x\sqrt{-G}(R-2 g_{ij}(\phi)\partial_\mu\phi^i \partial^\mu \phi^j-
  f_{ab}(\phi)F^a_{\mu \nu} F^{b \ \mu \nu} -{\textstyle{1 \over 2}} {\tilde f}_{ab}(\phi) F^a_{\mu \nu}
  F^b_{\rho \sigma} \epsilon^{\mu \nu \rho \sigma} ) ~.
  \label{actiongen}
\end{equation}
Assuming a spherically symmetric space-time metric ansatz of the  form,
\begin{eqnarray}
  ds^{2} & = & -a(r)^{2}dt^{2}+a(r)^{-2}dr^{2}+b(r)^{2}d\Omega^{2} ~, \label{metric2}
\end{eqnarray}
the gauge field equations are solved by \begin{equation}
  \label{fstrengthgen}
  F^a=f^{ab}(Q_{eb}-{\tilde f}_{bc}Q^c_m) {1\over b^2} dt\wedge dr + Q_m^a \sin \theta  d\theta \wedge d\phi ~,
\end{equation}
where $Q_m^a, Q_{ea}$ are constants that determine the magnetic and electric charges
carried by the gauge field $F^a$, and $f^{ab}$ is the inverse of $f_{ab}$. Defining an effective potential $V_\mathrm{eff}$ by,
\begin{equation}
  \label{defpotgen}
  V_\mathrm{eff}(\phi_i)=f^{ab}(Q_{ea}-{\tilde f}_{ac}Q^c_m)(Q_{eb}- {\tilde f}_{bd}Q^d_m)+f_{ab}Q^a_mQ^b_m ~,
\end{equation}
the equations of motion can be written
\begin{eqnarray}
(a^2 b^2)''&=& 2 ~, \label{1}\\
\frac{b''}{b}&=& - \phi'^2 ~, \label{2}\\
(a^2 b^2g_{ij}{\phi^j}')'&=& \frac{\partial_i V_\mathrm{eff}}{2 b^2} ~, \label{3} \\
-1 + a^2 b'^2 + \frac12 (a^2)' (b^2)' &=& - \frac{V_\mathrm{eff}}{b^2} + a^2 b^2 \phi'^2 ~, \label{constraint}
\end{eqnarray}
where $\phi'^2$ is short hand for $g_{ij}{\phi^i}'{\phi^j}'$.
Equation \eqref{constraint} is the Hamiltonian constraint that must be imposed on the field configurations that satisfy the equations of motion.
Using \eqref {2}, we express the constraint (\ref{constraint}) in a form that will be useful for our purposes:
\begin{equation}
\label{constraint2}
-\frac{(a^2 (b^2)')'}{2}= \frac{V_\mathrm{eff}}{b^2}-1 ~.
\end{equation}
It is easy to solve (\ref{1}) to obtain
\begin{equation}
a^2 b^2 = (r- r_+)(r- r_-)\label{5} ~.
\end{equation}
The non-extremal solutions we are interested in have distinct inner and outer horizons at $r_+$ and $r_-$, respectively.
Temperatures are obtained from the periodicity of the Euclidean time direction at the two horizons to be
\begin{eqnarray}
T_{\pm} = \frac{(a^2)'_\pm}{4\pi} ~,\label{6}
\end{eqnarray}
where  $+$ and $-$ subscripts denote the outer and inner horizons respectively.
We then employ (\ref{5}) to obtain the following expressions  for the temperatures:
\begin{equation}
4\pi T_\pm=(a^2)'_\pm = \pm \frac{(r_+ - r_-)}{b^2_\pm} ~. \label{7}
\end{equation}
This allows us to deduce the relation
\begin{equation}
b_+^{2} T_+ =  - b_-^2 T_- =\frac{r_+ -  r_-}{4\pi} = \frac{\Delta}{2\pi} ~, \label{8}
\end{equation}
where $\Delta = \tfrac{1}{2}({r_+  -  r_-})$ is a non-extremality parameter, which, as can be seen from (\ref{7}) and (\ref{8}), goes to zero in the
extremal case  and is directly proportional to the temperatures, $T_\pm$.
Evaluating (\ref{constraint}) at the horizons, we obtain,
\begin{equation}
-b_{\pm}^2 \pm \Delta (b_{\pm}^2)' = - V_{\pm}~,\label{ham}
\end{equation}
where $V_\pm$ denote the effective potential evaluated on the inner and
outer horizon.
Another useful relation is obtained by evaluating  (\ref{constraint})
at infinity giving \cite{attr5}:
\begin{equation}
  \label{eq:ham_inf}
  \Delta^2 = M^2+g_{ij}(\phi_\infty)\Sigma^i\Sigma^j-V_\mathrm{eff}(\phi_\infty)~,
\end{equation}
where we have used $2M=(r_++r_-)$\footnote{The equality $2M=(r_++r_-)$ follows from
  \eqref{5} if we take $b^2\rightarrow r^2$ asymptotically.} and $\Sigma^i$ is the tail
of the scalar field: $\phi^i=\phi^i_\infty+\Sigma^i/r+\ldots$ .

Taking the extremal limit of~(\ref{ham}), \textit{i.e.}, $\Delta\rightarrow0$, we
recover the relationship between the extremal entropy, $S_{\mathrm{ext}}$, the
extremal horizon radius, $b_{\mathrm{ext}}$ and the effective potential
evaluated at the horizon $V_\mathrm{eff}$ \cite{attr5,hep-th/0507096}:
\begin{equation}
  \label{eq:a_rel1}
S_{\mathrm{ext}}=\frac{1}{4}A_{\mathrm{ext}}= \pi b_{\mathrm{ext}}^2 = \pi V_\mathrm{eff}|_{\mathrm{Horizon}} ~.
\end{equation}
Furthermore, evaluating (\ref{3}) at the double horizon of an extremal
black hole, we find the values of the moduli are fixed at the horizon
by the attractor equation \cite{attr5,hep-th/0507096}
\begin{equation}
  \label{eq:a_rel2}
  \p_\phi V_\mathrm{eff}|_{\mathrm{Horizon}} =0 ~,
\end{equation}
which states that the effective potential at the horizon is independent of the asymptotic values of the moduli.
We see immediately from  (\ref{eq:a_rel1}) that
this in turn ensures the moduli independence of the entropy.
Now, (\ref{eq:a_rel1}), (\ref{eq:a_rel2}), which as discussed, essentially
encode the attractor mechanism,  can be written:
\begin{eqnarray}
\left[\frac{V_\mathrm{eff}(\phi)}{b^2}\,-\,1\right]_{\mathrm{Horizon}} &=&0 ~,\label{atr1}\\
\left.\frac{\partial_{\phi}V_\mathrm{eff}(\phi)}{b^2} \right|_{\mathrm{Horizon}}&=&0 ~.\label{atr2}
\end{eqnarray}
This form will be useful in the next section where we will see that they can be generalized to the non-extremal case
by averaging between the inner and outer horizons.

\section{Hot  attractor equations }
\label{sec:hot-attr-equat}
We consider non-extremal black hole solutions arising as solutions of
(\ref{actiongen}). They are characterized by gauge charges,
temperature and the asymptotic values of the moduli.  For fixed
temperature and charges there is a family of solutions with different
asymptotic moduli.  It turns out that by studying the inner and outer
horizons, we can generalize the attractor mechanism to the
finite temperature case.

We consider (\ref{8}) which, using the usual area-entropy relation,  can be written
\begin{equation}
  \label{eq:var1}
  S_+ T_+ + S_- T_- = 0 ~.
\end{equation}
Assuming that in general the first law holds at both horizons, keeping   angular momentum and gauge charges fixed, we have:
\begin{equation}
  \label{eq:first_law_constQ}
  dM=T_\pm dS_\pm ~,
\end{equation}
  so that
  \begin{equation}
    \label{eq:beta}
    T_\pm^{-1} = \left. \frac{\partial S_\pm}{\partial M} \right|_{J,Q} ~.
  \end{equation}
Now, consider the product of the entropies $S_+S_-$. Varying the mass subject to (\ref{eq:first_law_constQ}) and using (\ref{eq:var1}) gives
\begin{equation}
\label{eq:dSpSm}
 \left. \frac{\partial}{\partial M}(S_+S_-) \right|_{J,Q} = \frac{S_+T_++S_-T_-}{T_+T_-}=0 ~.
\end{equation}
In particular, since the product is independent of mass along the trajectory we have described, taking the extremal limit, $\Delta\rightarrow0$, yields
\begin{equation}
  \label{eq:mass_indep}
  S_+S_-=S_{\mathrm{ext}}^2 ~.
\end{equation}
Now by the attractor mechanism the extremal entropy, $S_{\mathrm{ext}}$ is
independent of the asymptotic moduli so $S_+S_-$ is as
well.\footnote{We are grateful to A.\ Castro for discussing this derivation with us.}

One possible flaw in this proof could be for the case where the extremal limit is singular --- for example, a null singularity --- so that we cannot invoke the attractor mechanism.
Moreover,~\eqref{eq:mass_indep} applies only when the Smarr relations are satisfied~\cite{cretc5}.
In Appendix~\ref{appc}, we show that in general
\begin{equation}
  \label{eq:tedious}
  T_+ dS_+ = T_- dS_-
\end{equation}
from which the invariance of $A_+A_-$ follows.

Given $a^2 b^2 = (r-r_+)(r-r_-)$ from \eqref{5}, we can define a Killing vector $\partial_\tau = a^2 b^2 \partial_r$ and write
\be
\int_c^\tau d\tau = \int_{r_0}^r \frac{dr}{(r-r_+)(r-r_-)} = \frac{1}{r_+-r_-} \int_{r_0}^r dr\ \left(\frac{1}{r-r_+} - \frac{1}{r - r_-} \right)~.
\ee
As there are poles, we express the coordinates in terms of three patches, which we define as follows
\begin{equation}
  \label{eq:tau_def}
\begin{array}{cccc}
\mbox{\textbf{Region 1:}} & r\in[0,r_-] ~, & \tau\in[0,\infty] ~, & \tau=\frac{1}{r_+-r_-} \log\left( \frac{1-\frac{r}{r_+}}{1-\frac{r}{r_-}} \right) ~, \\ 
\mbox{\textbf{Region 2:}} & r\in[r_-,r_+] ~, & \tau\in[\infty,-\infty] ~, & \tau=\frac{1}{r_+-r_-} \log\left( \frac{r_+-r}{r_--r} \right) ~, \\ 
\mbox{\textbf{Region 3:}} & r\in[r_+,\infty] ~, & \tau\in[-\infty,0] ~, & \tau=\frac{1}{r_+-r_-} \log\left( \frac{r-r_+}{r-r_-} \right) ~. 
\end{array}
\end{equation}

Integrating (\ref{3}) and (\ref{constraint2}) over Region 2, the region between the two horizons, and using
the fact that $a_\pm=0$ to evaluate the boundary terms, we find
averaged versions of the extremal attractor equations (\ref{atr1}),
(\ref{atr2}):
\begin{eqnarray}
\int^{r_+}_{r_-}dr\left( {\frac{V_\mathrm{eff}}{b^2} } -1\right) &=& \left[ -\frac{a^2 (b^2)'}{2} \right]^{r_+}_{r_-} = 0 ~, \label{eq:temp1} \\
\int^{r_+}_{r_-}dr\left( \frac{\partial_j V_\mathrm{eff}}{b^2}\right)&=& \left[ 2 a^2 b^2g_{ij}{\phi^j}' \right]^{r_+}_{r_-}= 0 \label{eq:temp2} ~.
\end{eqnarray}

\subsection{A small generalization: the scalar potential}
\label{sec:gauged}

We consider a generalization of (\ref{actiongen}) with an additional scalar potential $(-2V_g(\phi))$. Such a term would appear if we where to consider gauged 
supergravities for example. We also take a slightly more 
general ansatz:
\beq\label{metric}
ds^2= -a(r)^2 dt^2+a(r)^{-2}dr^2+ b(r)^2d\Omega^2_{k}~,
\eeq
where the label $k$ in (\ref{metric}) denotes  the assumed metric of the transverse spacial foliation with:
\begin{equation}
  \label{eq:trans_metric}
  d\Omega_k^2=
  \left\{
    \begin{array}{ll}
       d\theta^2 + \sin^2\theta d\phi^2 & k=1 \\
       dx^2+dy^2 & k=0 \\
       d\theta^2 + \sinh^2\theta d\phi^2 & k=-1 \\
    \end{array}
  \right. ~.
\end{equation} 
The equations of motion (\ref{1}), (\ref{3}), (\ref{constraint}) become  
\begin{eqnarray}
(a^2 b^2)''&=& 2(\,k\,-2\,V_g b^2\,) ~, \label{1V}\\
(a^2 b^2g_{ij}{\phi^j}')'&=&  \tfrac{1}{2}\left(\frac{\partial_i V_\mathrm{eff}}{ b^2} +\partial_i V_g b^2\right)~, \label{3V} \\
-k + a^2 b'^2 + \frac12 (a^2)' (b^2)' &=& - \frac{V_\mathrm{eff}}{b^2} - b^2 V_g + a^2 b^2 \phi'^2 ~, \label{constraintV}
\end{eqnarray}
while (\ref{2}) remains the same. 
Using \eqref {1V} and \eqref{2}, the constraint (\ref{constraintV}) can once again be written in terms of a total derivative :
\begin{equation}
\label{constraint2V}
-\frac{(a^2 (b^2)')'}{2}= \frac{V_\mathrm{eff}}{b^2}+b^2 V_g -k ~.
\end{equation}
In a similar fashion to the preceding section  integration of (\ref{3V}) and (\ref{constraint2V}) over Region 2 leads to the following averaged attractor equations:
\begin{eqnarray}
\int^{r_+}_{r_-}dr\left( {\frac{V_\mathrm{eff}}{b^2} }+b^2V_g -k\right) &=& \left[ -\frac{a^2 (b^2)'}{2} \right]^{r_+}_{r_-} = 0 ~, \label{eq:temp1V} \\
\int^{r_+}_{r_-}dr\left( \frac{\partial_i V_\mathrm{eff}}{b^2}+ b^2 \partial_i V_g \right)&=& \left[ 2 a^2 b^2g_{ij}{\phi^j}' \right]^{r_+}_{r_-}= 0 \label{eq:temp2V} ~.
\end{eqnarray}
Unfortunately our results for the invariance of horizon area products do not carry over to this case since the right hand side of \eqref{1V} is not constant 
in general.  For interesting results regarding area product formulas in gauged supergravity see for example \cite{cgp,Toldo:2012ec}.
\section{Moduli space mysteries}
\label{sec:mod}

For extremal black holes, the attractor mechanism ensures that the
moduli take on fixed values on the horizon such that the effective
potential is minimized ensuring that entropy does not depend on the
asymptotic moduli.  As we have seen from the preceding sections, some
simple geometric properties (\ref{1}), (\ref{6}) and the first law
ensure that, for non-extremal black holes, the product of entropies is
independent of the asymptotic moduli. However, the
entropies on their own do generically depend on these values~\cite{hep-th/0507096,hep-th/061114}. Somehow
the nature of the flow between the inner and outer horizons ensures
that there is a cancellation. While we have found averaged
versions of the attractor equations, it is not clear how they
explicitly relate to the collective invariance of the entropies and the behavior of the flow.

We have examined some explicit solutions to try gain some insight to this question.
Our observations, based on an examination of the solutions, are summarized in Table~\ref{tab:sols} with further details in Appendix \ref{sec:exact}.
Scanning Table~\ref{tab:sols}, there does not seem to be any
clear universal features when we compare the values of the moduli or the effective potential
on the two horizons. We will see below that some suggestive results
emerge from a perturbation analysis.
\begin{table}[htb]
  \centering
\begin{tabular}{|l|l|c|c|c|c|c|}
\hline
Case & $V_\mathrm{eff}$  & $\phi_\pm$ & $V_\pm$ & $\lambda^2$ &$\nu$ \\ \hline\hline
1& $\alpha_1=-\alpha_2=2$ &  $\phi_0=\tfrac{1}{2}(\phi_+ + \phi_-)$ & $V_+=V_-$ & 2 &1 \\ \hline
$1'$& $\alpha_1\alpha_2=-4$ & $\phi_0=\tfrac{1}{2}(\phi_+ + \phi_-)$ & $V_+\neq V_-$ generically & 2 &1 \\ \hline
2& $\alpha_1=-\alpha_2=2\sqrt3$ & $\phi_-=\phi_+$ & $V_+=V_-$& 6 &2\\ \hline
\end{tabular}

  \caption{Summary of features of moduli flow between the horizons for exact solutions.
All the examples considered have an effective potential of the form $V_\mathrm{eff}=e^{\alpha_1}(Q_1)^2+e^{\alpha_2}(Q_2)^2$.
Here, $\phi_0$ is the attractor value of the scalar which minimizes $V_\mathrm{eff}$.
The two parameters $\lambda^2=\frac{\p_\phi^2V_\mathrm{eff}|_{\phi=\phi_0}}{2V_\mathrm{eff}|_{\phi=\phi_0}}$ and  $\nu=\tfrac{1}{2}(\sqrt{1+4\lambda^2}-1)$
arise  in the perturbation analysis as discussed in Appendix~\ref{sec:pert}.}
  \label{tab:sols}
\end{table}
 \begin{figure}[htb]
  \centering
  \includegraphics[width=0.86\linewidth]{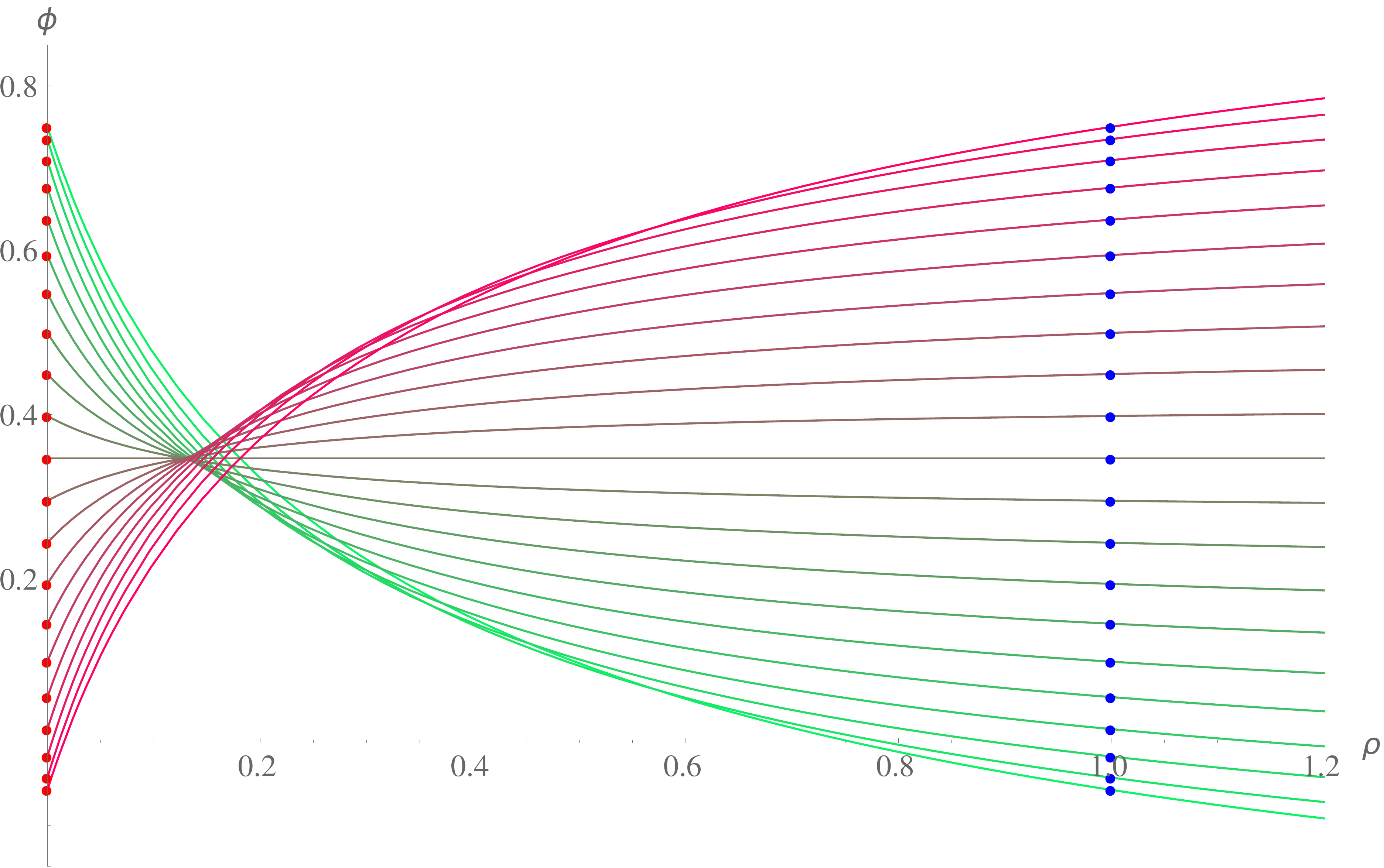}
 \includegraphics[width=0.13\linewidth]{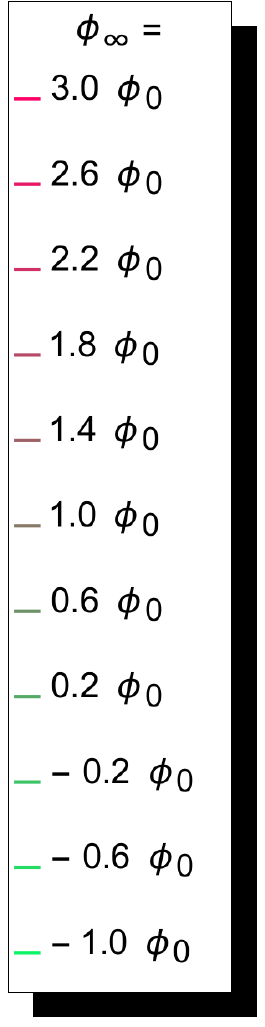}
  \caption{Modulus flow for Case 1 ($V_\mathrm{eff}=e^{2\phi}(Q_1)^2+e^{-2\phi}(Q_2)^2$) between the horizons for various asymptotic values $\phi_\infty$ with $Q_1=1$, $Q_2=2$ and $M=3$.
For these parameters, $\phi_{0}=\frac{1}{2}\ln|Q_2/Q_1|\approx 0.35$. We use a scaled
radial coordinate $\rho$ defined by $r=(r_+-r_-)\rho + r_-$ so that outer and inner horizons are at $\rho=1$ and $0$ respectively.}
  \label{fig:flow1}
\end{figure}

\begin{figure}[htb]
  \centering
  \includegraphics[width=0.86\linewidth]{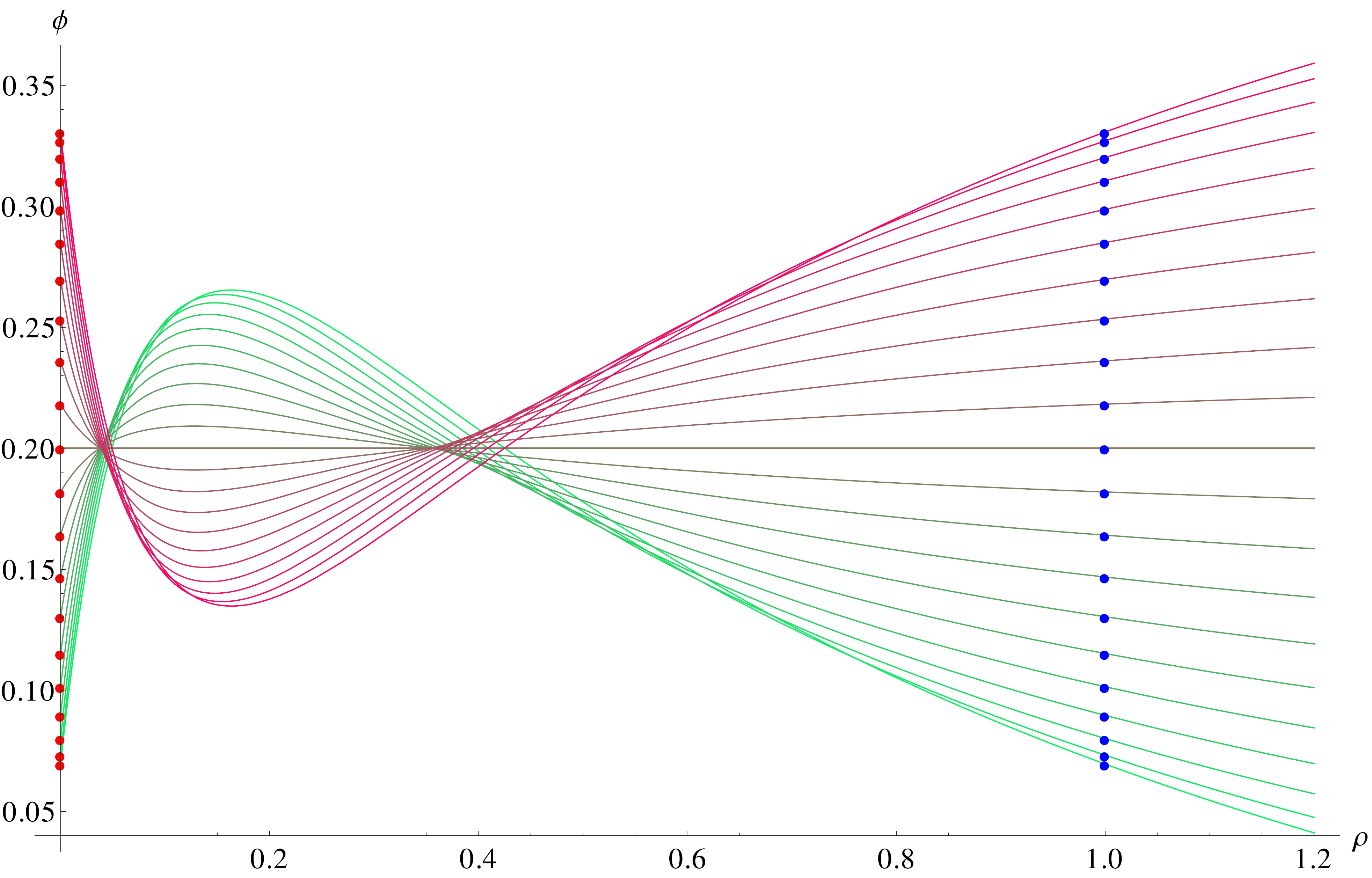}
 \includegraphics[width=0.13\linewidth]{legend.pdf}
  \caption{Modulus flow for Case 2 ($V_\mathrm{eff}=e^{2\sqrt3\phi}(Q_1)^2+e^{-2sqrt3\phi}(Q_2)^2$) between the horizons for various asymptotic values $\phi_\infty$ with $Q_1=1$, $Q_2=2$ and $M=3$.
For these parameters, $\phi_{0}=\frac{1}{2\sqrt3}\ln|Q_2/Q_1|\approx 0.20$. Once again we use the scaled
radial coordinate $\rho$ so that outer and inner horizons are at $\rho=1$ and $0$ respectively.}
  \label{fig:flow2_1}
\end{figure}

\begin{figure}[htb]
  \centering
  \includegraphics[width=0.86\linewidth]{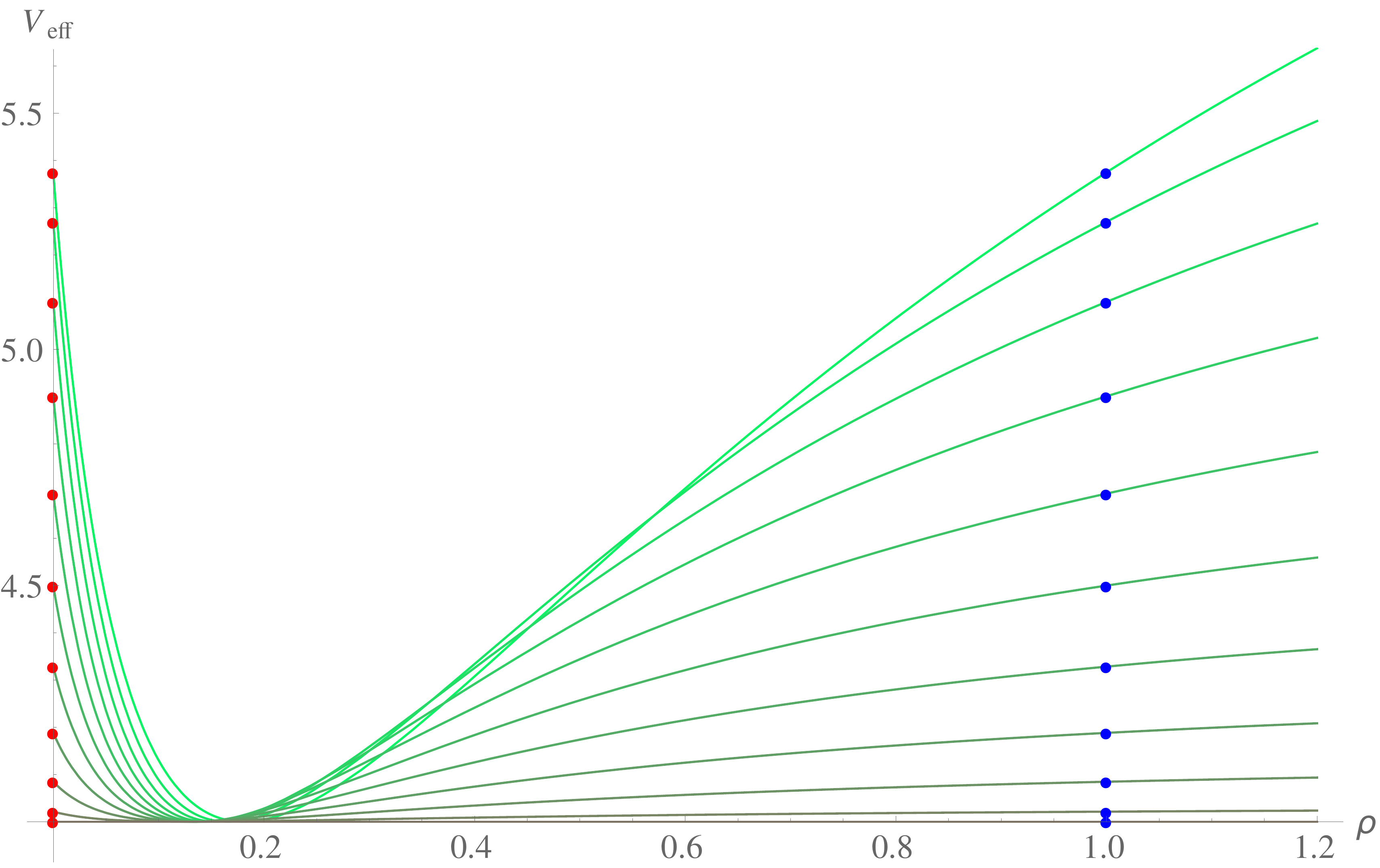}
 \includegraphics[width=0.13\linewidth]{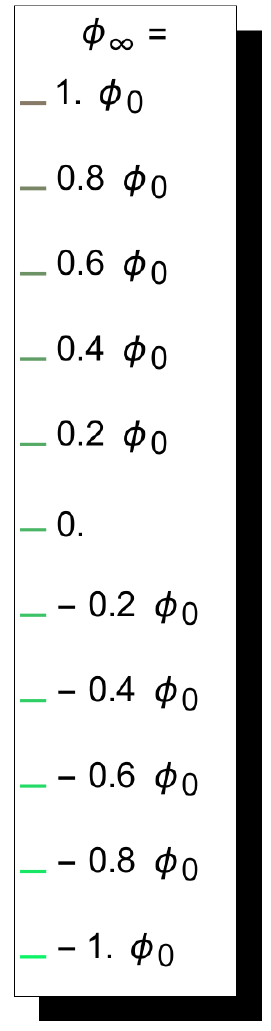}
  \caption{Behavior of $V_\mathrm{eff}$ for Case 1 between the horizons}
  \label{fig:flow2}
\end{figure}
\begin{figure}[htb]
  \centering
  \includegraphics[width=0.86\linewidth]{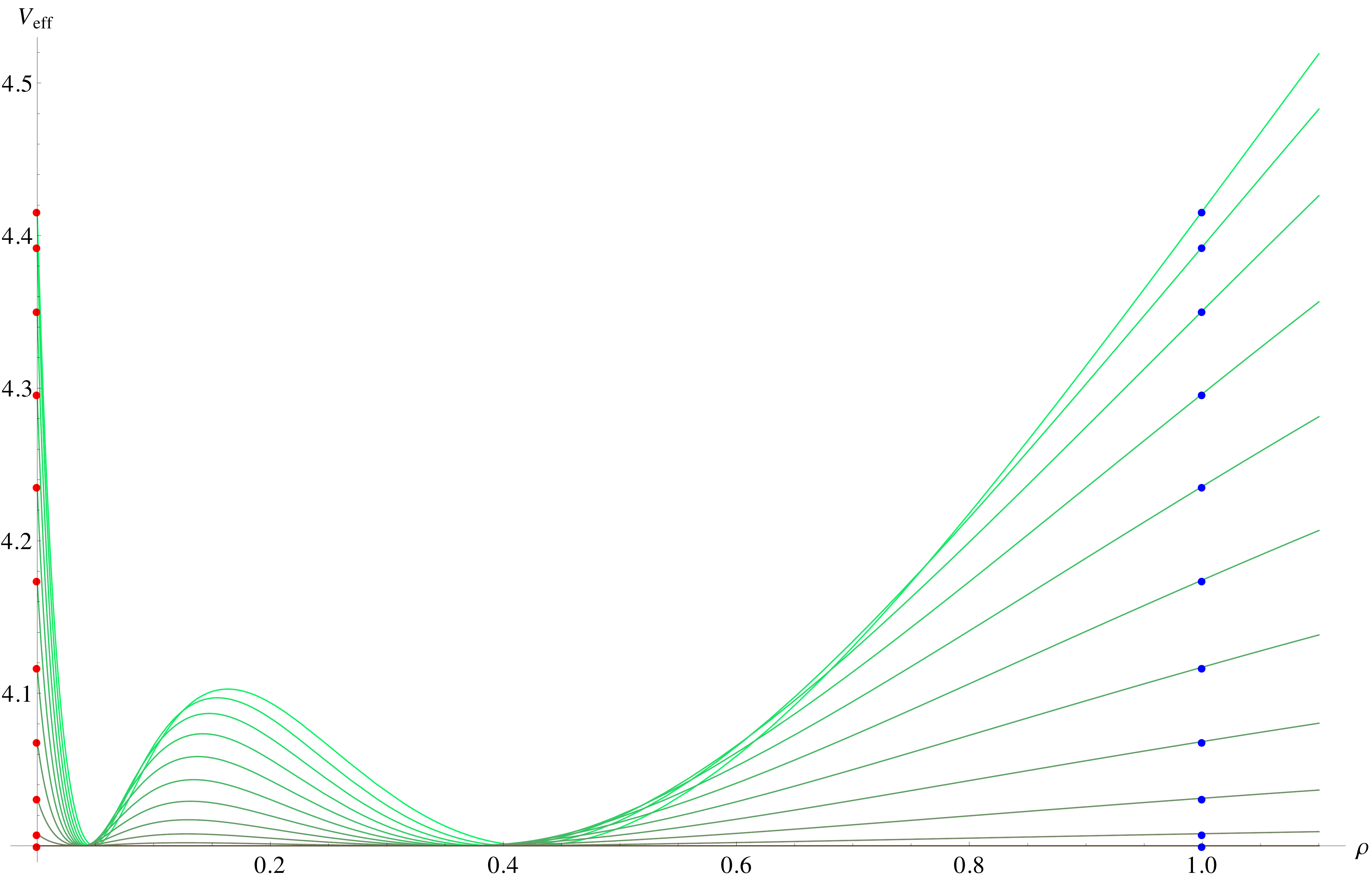}
 \includegraphics[width=0.13\linewidth]{legend2.pdf}
  \caption{Behavior of $V_\mathrm{eff}$ for Case 2 between the horizons}
  \label{fig:flow2_2}
\end{figure}

We have plotted various flows in Figures~
\ref{fig:flow1}--\ref{fig:flow2_2},
to illustrate the features of the solutions. In the plots we
have kept $M$ and $Q_i$ constant while varying $\phi_\infty$.   Figure \ref{fig:flow1} shows Case
1, in which, as shown in Table \ref{tab:sols}, the modulus averaged
over the horizons give the attractor value.
We see that the non-constant flows intersect the attractor value
once. Figure \ref{fig:flow2_1}  shows Case 2,
in which the modulus has the same value on both horizons. We see that
the flows intersect the attractor value twice.
The plots do not intersect the attractor value at a common point. Plots of the effective potential, Figures
\ref{fig:flow2}--\ref{fig:flow2_2} do not unfortunately seem to yield
any additional insight --- we simply verify the effective potentials
attain their minimum when the modulus passes through the attractor
value as expected.

We also tried to extend the perturbation strategy of
\cite{hep-th/0507096, hep-th/061114} to the non-extremal case
now considering both horizons.  Even in the
non-extremal case, we can solve the scalar equation of motion
(\ref{3}) by fixing them to their attractor value, $\phi=\phi_0$,
\textit{i.e.}, such that $\partial_\phi V_\mathrm{eff}=0$.  This leaves us
simply with a Reissner--Nordstr\"om black hole metric. We
observe from \eqref{eq:ham_inf} that with constant scalars,
\begin{equation}
  \label{eq:rprm}
r_+r_-= M^2-\Delta^2=V_\mathrm{eff}(\phi_0)~.
\end{equation}
Perturbing about the constant scalar solutions (see Appendix
\ref{sec:pert}) we found that unless
\[\nu=\frac{1}{2}\left(\sqrt{1+4\left(\frac{\p_\phi^2V_\mathrm{eff}|_{\phi=\phi_0}}{2V_\mathrm{eff}|_{\phi=\phi_0}}\right)^2}-1\right)~,\]
is an integer,  the first order scalar perturbation
blows up on at least one of the horizons. It is not clear whether this
divergence relates to the instability of the inner horizon or the
possibility that we can only find non-trivial flows when $\nu\in\mathbb{Z}$.
The exact solutions we considered correspond to $\nu=1,2$. We found that
the first order perturbation goes like the Legendre polynomial
$P_\nu$, so that for the limited sample of exact solutions considered, the number of zeros of the perturbation correlates with the
number of times the flow passes through the attractor value between
the horizons. We {\it speculate} that
\begin{itemize}
\item for single scalar field flows in general, $\nu$
corresponds to the number of times a non-constant  flow passes through the
attractor point,
\item $\delta\phi_+=(-)^\nu\delta\phi_-$ (where $\phi=\phi_0+\delta\phi$),
 \end{itemize}
  which 
    is consistent with Table
  \ref{tab:sols} and the first order perturbation analysis. 
On the other hand, it may be that the averaged attractor equations (\ref{atr1}), (\ref{atr2}) are the most general
statements we can make.

\section{CFT interpretation}
\label{sec:cft}
The microscopic interpretation of general extremal black hole thermodynamics relies on the fact that  an extremal solution admits a near-horizon AdS$_2$ geometry with the scalars fixed at their attractor values.   If, at  a certain point in moduli space, this can be lifted to a BTZ black hole with an AdS$_3$ factor, one can then think of this as a state in the holographically dual CFT$_2$ and relate the relevant bulk thermodynamic quantities to excitation numbers and central charges in the CFT and thereby acquire a microscopic perspective. In order to attain a CFT interpretation, we therefore attempt to use the symmetries of the equations of motion to transform any given asymptotically flat fixed scalar black hole background to an asymptotically AdS$_2 \times S^2$ background with the near-horizon geometry remaining unchanged. This implies that any thermodynamic interpretation we may wish to impose upon black hole quantities at the horizons remain unaffected.

We find inspiration in the work of~\cite{cl1,cl2,cg,bdjm}, which maps black hole spacetimes with flat asymptopia to black hole spacetimes with AdS asymptopia by a judicious modification of the warp factor.
Such subtracted geometries, which are the consequence of a Harrison transformation, transplant the black hole from one geometry to another while leaving the thermodynamic properties of the solution unchanged.
The conformal symmetry of the wave equation motivates the choice of warp factor.

Consider the transformation
\begin{equation}
a \rightarrow \Lambda (r) \, a ~, \qquad
b \rightarrow \frac{b}{\Lambda(r)} ~.
\end{equation}
We can show that this transformation is a symmetry of the equations of motion  for a fixed scalar background if and only if $\Lambda (r) = \frac{r}{\sqrt{r_+ r_-}}$. Hence, one can design a new black hole background with fixed scalar fields under this transformation. The resulting background has a metric given by
\begin{equation}
ds^2= - \frac{(r-r_-)(r-r_+)}{\ell^2} dt^2 +\frac{\ell^2}{(r-r_+)(r-r_-)} dr^2  + \ell^2 d\Omega_2^2 ~,
\end{equation}
which is an asymptotically AdS$_2 \times S^2$ space with the AdS radius given by $\ell^2 = V_\mathrm{eff}(\phi_0)$. We can then uplift an axion-free solution of this form to a BTZ black hole in AdS$_3 \times S^2$ with the metric
\begin{equation}
ds^2=    - \frac{(r-r_-)(r-r_+)}{\ell^2} dt^2 +\frac{\ell^2}{(r-r_+)(r-r_-)} dr^2 + r^2 (dy + \frac{r_+ r_-}{\ell r} dt)^2  + \ell^2 d\Omega_2^2 ~.
\end{equation}
We observe that the horizon coordinates of the BTZ black hole are the same as the original black hole in four dimensions\footnote{The AdS$_3$ part is not given in the usual form of a BTZ black hole,
but since BTZ is the unique AdS$_3$ black hole, the two must be
related by a coordinate transformation.}.
We can express the mass and angular momentum in terms of the Virasoro generators $L_0$ and $\overline{L}_0$ of CFT$_2$ in the usual manner:\footnote{For convenience, we work in units where $8 G_3 = 1$.}
\begin{eqnarray}
M = \frac{r_+^2 + r_-^2}{\ell^2} & \qquad \Longrightarrow \qquad & M\ell = L_0 + \overline{L}_0 ~,\\
J = \frac{2 r_+ r_-}{\ell} & \qquad \Longrightarrow \qquad & J = L_0 - \overline{L}_0 ~.
\end{eqnarray}

Defining the parameter $\tau = y + it_E$, the partition function of the two-dimensional CFT is
\be
Z = \mathrm{Tr}\, q^{L_0} \bar{q}^{\overline{L}_0} ~, \qquad q=e^{2\pi i \tau} ~.
\ee
The Peccei--Quinn transformation in the $SL(2,\mathbb{Z})$ modular group of the CFT  shifts the real part of the period variable $\tau$ by an integer quantity.
As this acts as the chemical potential for the generator of spatial translations along the asymptotic circle defined by $L_0 - \overline{L}_0$.
The eigenvalues correspond to momentum on this circle.
The generator acts as a symmetry of the two-dimensional partition function only when the eigenvalues of $L_0 - \overline{L}_0$ are quantized.
This is consistent with the assignment of this quantity to the angular momentum charge $J$.
The generator of translations on the time circle of the boundary torus of AdS is $L_0 + \bar{L}_0$.
In changing the mass of a black hole at fixed quantum number, we change the temperature and therefore the periodicity of this Euclidean time circle.
Under this process, \textit{only} the eigenvalue of translations along this circle changes  with the mass.
This corroborates the identification of the mass with the generator $L_0 + \overline{L}_0$.

As we consider different eigenvalues of $L_0 + \overline{L}_0$, the eigenvalue on the orthogonal circle determined by $L_0 - \overline{L}_0$ remains fixed.
Consequently, along the orbit described by $L_0 + \overline{L}_0$, the product $r_+ r_-$ is independent of the non-extremality parameter and is quantized in terms of the charges.
At extremality, we know that $\overline{L}_0 = 0$, and hence $M\ell =J$.
Here, $r_+ r_- = r_{\mathrm{ext}}^2$ unambiguously and this remains constant as the mass changes.
This calculation supplies a microscopic justification for the independence of the product of the horizon areas as the non-extremality parameter is varied.

\section{Discussion and conclusions}
\label{sec:disc}
The attractor mechanism relies on extremality, not supersymmetry~\cite{hep-th/0507096}.
It establishes that extremal black hole solutions in ${\cal N}=2$ four-dimensional gauged supergravity backgrounds have a horizon area that is independent of the asymptotic moduli.
The horizon acts as an attractor point in the moduli space towards which all the moduli flows converge.
We explore equivalent properties of the horizon and the moduli space in the non-extremal case.
We find that the product of the inner and outer horizon areas is moduli independent and write down conditions on the moduli space flows for non-extremal backgrounds.

These results could be better understood by looking at how the near-horizon geometry is analyzed in the extremal case. Here, we zoom in on  the geometry near the extremal horizon and analyze the moduli in this region which is AdS$_2 \times S^2$ and by itself is a complete solution of the equations of motion. The various symmetries of this background constrain the scalar moduli to be constant and hence the one-dimensional Lagrangian density becomes purely a function of the constant effective potential. Extremizing the potential with respect to the scalar moduli give their extremum values at the horizon, and the extremum value of the potential is proportional to the horizon area.  Recapitulating \eqref{atr1}, \eqref{atr2}, this can be written as
\begin{eqnarray}
\left<\frac{V_\mathrm{eff}(\phi)}{b^2}\,-\,1\right>_{\mathrm{AdS}_2 \times S^2}&=&0\,,\\
\left <\frac{\partial_{\phi^i} V_\mathrm{eff} (\phi)}{b^2}\right>_{\mathrm{AdS}_2\times S^2} &=&0 ~,
\end{eqnarray}
where the angle brackets emphasize the geometry of the near-horizon region.

In the non-extremal case, the role of the near-horizon geometry is fulfilled by Region 2, the region between the two horizons.
If the angle brackets above are interpreted to mean the average value in Region 2, the equations are the same.
Thus, our analysis has taken an initial step to extending studies of moduli
flow from extremal backgrounds to the non-extremal ones. Every
statement that we can make about the moduli space in extremal
backgrounds is encoded in the attractor mechanism, and we can now make
corresponding statements for non-extremal black holes, as displayed in Table \ref{table:comp}.
\begin{table}[htb]
  \centering
\begin{tabular}{|c|c|}
\hline
Extremal BHs & Non-Extremal BHs\\ \hline\hline
Moduli-independent area & Moduli-independent product of areas \\ \hline
$\left<\frac{V_\mathrm{eff}(\phi)}{b^2}\,-\,1\right>_{\mathrm{AdS}_2 \times S^2}=0$&$\left<\frac{V_\mathrm{eff}(\phi)}{b^2}\,-\,1\right>_\mathrm{Region\ 2}=0$\\ \hline
$\left <\frac{\partial_{\phi^i} V_\mathrm{eff} (\phi)}{b^2}\right>_{\mathrm{AdS}_2\times S^2} =0$&$\left <\frac{\partial_{\phi^i} V_\mathrm{eff} (\phi)}{b^2}\right>_\mathrm{Region\ 2} =0$\\ \hline
\end{tabular}
\caption{Comparison of hot and cold attractors}
\label{table:comp}
\end{table}

The attractor mechanism for the extremal black holes fixes the moduli
at the horizon purely in terms of the charges, thereby lowering the
degrees of the freedom of the system to half of the number required to
determine a generic solution. This allows us to formulate first order
equations of motion, which can be solved to obtain extremal solutions.  It has been
known for sometime that in certain cases, one can find first order
equations describing non-extremal flows, \cite{Lu:2003iv,hep-th/0612308}.
Although there has been a lot of
progress constructing such flows --- see, for example,
\cite{Andrianopoli:2007gt,Cardoso:2008gm,Perz:2008kh,Galli:2011fq,1108.0296,Meessen:2012su,Gnecchi:2014cqa,Goldstein:2014qha}
--- a general understanding remains elusive.  The existence of an
``attractor'' mechanism for a generic non-extremal solution may yield
deeper understanding of first order flows and help in finding new
solutions.

The recent article~\cite{Martinec} argues that the $S_+ S_- = S_\mathrm{ext}^2$ relation supports the microstate picture of black holes. (See~\cite{math1, math2, skta, benawarner} for reviews.) According to~\cite{Martinec}, the inner and outer horizons delimit the capped region in which long string degrees of freedom are localized. In light of these observations and our complementary analysis of the geometric mean formula using the attractor mechanism, non-extremal solitonic solutions in supergravity~\cite{me, bgrw, ccdm, dgr, bnw, nie} should be investigated further. This may be a crucial laboratory for describing the thermodynamics of inner horizons.

Finally, although our results only apply fully to Lagrangians of the form \eqref{actiongen} we 
believe it should be possible to generalise them to higher derivative theories and gauged supergravities. 

\section*{Acknowledgements} 
We are grateful to Alejandra Castro for the elegant proof that $S_+
S_- = S_\mathrm{ext}^2$ in Section~3.
We thank Costas Bachas, Paolo Benincasa, Roberto Emparan, Stefan Hollands, Finn Larsen, Niels Obers, Chiara Toldo, Jan Troost, and \'Alvaro V\'eliz-Osorio for discussions.
The work of KG is supported in part
by the National Research Foundation. 
The research of VJ and SN is supported by the South African Research
Chairs Initiative of the Department of Science and Technology and the
National Research Foundation.  VJ is grateful to the string group at
Queen Mary, University of London for kindly hosting him during the
concluding stages of this work.

\appendix

\section{Exact solution}
\label{sec:exact}

In this appendix, we  review some features of known exact solutions and derive some
properties summarized in Table \ref{tab:sols}.

\subsection{Case $1$}

Consider the effective potential,
$V_\mathrm{eff}=e^{2\phi}(Q_1)^2+e^{-2\phi}(Q_2)^2$ whose solutions can be written \cite{hep-th/9205027}
\begin{eqnarray}
  \label{eq:nonex}
  \exp(2\phi) & = & e^{2\phi_{\infty}}\frac{(r+\Sigma)}{(r-\Sigma)}\nonumber~, \\
  a^{2} & = & \frac{(r-r_{+})(r-r_{-})}{(r^{2}-\Sigma^{2})}\label{eq:alpha2_solution}~,\\
  b^{2} & = & (r^{2}-\Sigma^{2})~,\nonumber
\end{eqnarray}
where
\begin{equation}
\label{rhne}
  r_{\pm}=M\pm \Delta,\quad \bar Q_1=e^{\phi_\infty}Q_1,\quad \bar Q_2=e^{-\phi_\infty}Q_2~.
\end{equation}
In this case, the Hamiltonian constraint \eqref{eq:ham_inf} becomes
\begin{equation}
  \Sigma^{2}+M^{2}-\bar{Q}_{1}^{2}-\bar Q_{2}^{2}=\Delta^{2}~.
\end{equation}
The scalar charge, $\Sigma$, is not an independent
parameter --  it is given by
\begin{equation}
\label{sigma}
  \Sigma=\frac{\bar{Q}_2^{2}-\bar{Q}_1^{2}}{2M}~.
\end{equation}
Now, the attractor value, $\phi_0$, which extremizes the effective potential, is 
\begin{equation}
  \label{eq:atr}
  e^{2\phi_0}=|Q_2/Q_1|~,
\end{equation}
and one can check using (\ref{eq:nonex}), (\ref{rhne}), (\ref{sigma}) that
\begin{equation}
  \label{eq:atr2}
  e^{2\phi_+}e^{2\phi_-}=e^{4\phi_\infty}\frac{(M+\Sigma)^2-r_0^2}{(M-\Sigma)^2-r_0^2}=Q_2^2/Q_1^2=e^{4\phi_0}~,
\end{equation}
so that the scalar averaged on both horizons gives the attractor value:
\begin{equation}
  \label{eq:atr3}
  \phi_0=\tfrac{1}{2}(\phi_+ + \phi_-)~.
\end{equation}
From (\ref{eq:atr}), (\ref{eq:atr3}) we can see that
\begin{eqnarray}
  \label{eq:Vpm}
  V_{+}&=&e^{4\phi_0}e^{-2\phi_-}Q_1^2+e^{-4\phi_0}e^{2\phi_-}Q_2^2=V_{-}\\
  (\p_\phi V_\mathrm{eff})_+&=& 2e^{4\phi_0}e^{-2\phi_-}Q_1^2-2e^{-4\phi_0}e^{2\phi_-}Q_2^2= -(\p_\phi V_\mathrm{eff})_-
\end{eqnarray}
As expected,
\begin{eqnarray}
  \label{eq:bbb}
  b_+^2b_-^2&=&(r_+^2-\Sigma^2)(r_-^2-\Sigma^2)\\
  &=& (M^2+r_0^2-\Sigma^2)-4M^2r_0^2\\
  &=& (\bar Q_1^2+\bar Q_2^2)^2 - (\bar Q_1^2-\bar Q_2^2)^2\\
  &=& 4 Q_1^2 Q_2^2 = b_0^4~,
\end{eqnarray}
is independent of $\phi_\infty$.
\subsection{Case ${1'}$}
Consider the effective potential,
$V_\mathrm{eff}=e^{\alpha\phi}(Q_1)^2+e^{-\frac{4}{\alpha}\phi}(Q_2)^2$ whose solutions can be written  \cite{Gibbons:1987ps,hep-th/0507096}
\begin{eqnarray}\label{1b_sol}
e^{(\frac{\alpha}{2}+\frac{2}{\alpha})\phi}&=&\left(\frac{2}{|\alpha|}\right)\left(\frac{|Q_2|F_2}{|Q_1|F_1}\right)~,\\
a^2&=& \Delta^2(Q_1F_1)^{-\frac{8/\alpha}{\alpha+4/\alpha}}(Q_2F_2)^{-\frac{2\alpha}{\alpha+4/\alpha}}/\spadesuit~,
\end{eqnarray}
where
\begin{equation}
  F_i = \sinh(\Delta\tau-\log(e_i))~,\qquad
\spadesuit=(4/\alpha^2)^{\alpha/(\alpha+4/\alpha)}+(4/\alpha^2)^{4/\alpha/(\alpha+4/\alpha)}~,
\end{equation}
and $\tau$ is given by \eqref{eq:tau_def},
so that
\begin{equation}
  b^2=\spadesuit(\mu_1r-\nu_1)^{8/(4+\alpha^2)}(\mu_2r-\nu_2)^{2\alpha^2/(4+\alpha^2)}/\Delta^2~,
\end{equation}
where
\begin{eqnarray}
  \mu_i &=&{Q_i(e_i^2-1)}/({2e_i})~,\\
  \nu_i &=& (r_+-r_-e_i^2)/(2e_i)~,
\end{eqnarray}
and the $e_i$ are given by \cite{hep-th/061114}
\begin{equation}
  \label{eq:e_def}
  \tfrac{1}{2}\left(e_i -e_i^{-1}\right)=2\Delta(\alpha_i^2+4){(\bar Q_i)}^{-1}~,
\end{equation}
with $(\alpha_1,\alpha_2)=(\alpha,-4/\alpha)$ and $\bar Q_i^2=e^{\alpha_i\phi_\infty}Q_i^2$.

On the inner and outer horizons, we obtain,
\begin{eqnarray}
  e^{(\frac{\alpha}{2}+\frac{2}{\alpha})\phi_+}&=& \frac{2e_2Q_2}{e_1Q_1\alpha}~,\\
    e^{(\frac{\alpha}{2}+\frac{2}{\alpha})\phi_-}&=& \frac{2e_1Q_2}{e_2Q_1\alpha}~,
\end{eqnarray}
so
\begin{equation}
  e^{(\frac{\alpha}{2}+\frac{2}{\alpha})\phi_+}e^{(\frac{\alpha}{2}+\frac{2}{\alpha})\phi_-}
  =
  \frac{4Q_2^2}{Q_1^2\alpha^2}= e^{2(\frac{\alpha}{2}+\frac{2}{\alpha})\phi_0}~,
\end{equation}
and we again have
that the scalar averaged on both horizons gives the attractor value:
\begin{equation}
  \label{eq:atr3b}
  \phi_0=\tfrac{1}{2}(\phi_+ + \phi_-)~.
\end{equation}
We also have that
\begin{eqnarray}
  b_+^2&=&\spadesuit(-e_1 Q_1)^{8/(4+\alpha^2)}(-e_2Q_2)^{2\alpha^2/(4+\alpha^2)}~,\\
  b_-^2&=&\spadesuit(-Q_1/e_1)^{8/(4+\alpha^2)}(-Q_2/e_2)^{2\alpha^2/(4+\alpha^2)}~.
\end{eqnarray}
So that as expected
\begin{equation}
  b_+^2b_-^2=b_0^4~.
\end{equation}
However, unlike the previous case, one can show numerically that one does not  generically have  $V_+= V_-$.

\subsection{Case $2$}

Consider the effective potential,
$V_\mathrm{eff}=e^{2\sqrt{3}\phi}(Q_1)^2+e^{-2\sqrt{3}\phi}(Q_2)^2$ whose solutions \cite{Leutwyler,Dobiasch:1981vh,Chodos,Pollard} can be written \cite{hep-th/0507096}
\begin{eqnarray}\label{toda0}
e^{\frac{4}{\sqrt{3}}\phi}&=&\left(\frac{Q_2}{Q_1}\right)^{2/3}\frac{e^{-q_2}}{e^{-q1}}~,\\
a^2&=& \frac{e^{\frac{1}{2}(q_1+q_2)}}{\sqrt{16 Q_1 Q_2}}\label{toda1}~,
\end{eqnarray}
where the $q_i$ satisfy the two particle toda equation
\begin{eqnarray}
\label{toda2 }
  \ddot q_1&=&e^{2q_1-q_2}~,\\
\ddot q_2&=&e^{2q_2-q_1}~,
\end{eqnarray}
and the dot denotes derivatives with respect to $\tau$ which is
defined by \eqref{eq:tau_def}.
Solutions to (\ref{toda2 }) can be written
\begin{eqnarray}
  \label{eq:toda_sol}
  e^{-q_1}&=&f_1e^{m_1\tau}+f_2e^{m_2\tau}+f_3e^{-(m_1+m_2)\tau}~,\\
    e^{-q_2}&=&f_4e^{-m_2\tau}+f_5e^{-m_1\tau}+f_6e^{(m_1+m_2)\tau}~,
  \end{eqnarray}
where
\begin{eqnarray}
  \label{eq:const1}
  f_3&=&\frac{1}{f_1f_2(m_1-m_2)^2(2m_1+m_2)^2(m_1+2m_2)^2}~, \\
  f_4&=& \frac{1}{f_2(m_1-m_2)^2(m_1+2m_2)^2}~, \label{eq:const2} \\
  f_5&=& \frac{1}{f_1(m_1-m_2)^2(2m_1+m_2)^2}~, \label{eq:const3}\\
  f_6&=& -f_1f_2(m_1-m_2)^2~, \label{eq:const4}
\end{eqnarray}
and matching with the known solutions (or imposing $\phi$ finite on the horizon and finite horizon area)
fixes $m_1=-m_2 $ and $m_1=\pm(r_+-r_-)$ -- without a loss of
generality we may take $m_1=(r_+-r_-)=2\Delta$. The constants $f_{1,2}$ depend on $Q_{1,2}$ and $\phi_\infty$ in a complicated way. Using
\eqref{eq:tau_def} we get
\begin{eqnarray}
\label{toda_sol2a}
  e^{-q1}&=&\frac{f_1(r-r_+)^2+f_2(r-r_-)^2+f_3(r-r_+)(r-r_-)}{(r-r_+)(r-r_-)}~,\\
    e^{-q2}&=&\frac{f_4(r-r_+)^2+f_5(r-r_-)^2+f_6(r-r_+)(r-r_-)}{(r-r_+)(r-r_-)}\label{toda_sol2b}~.
\end{eqnarray}
Now using (\ref{5}), (\ref{toda0}), (\ref{toda_sol2a}), (\ref{toda_sol2b}), we get
\begin{eqnarray}
  \label{eq:toda_horizon}
  e^{\frac{4}{\sqrt{3}}\phi_+}&=&\left(\frac{Q_2}{Q_1}\right)^{2/3}\frac{f_5}{f_2}=-\left(\frac{Q_2}{Q_1}\right)^{2/3}\frac{1}{64 f_1f_2\Delta^4}~,\\
  e^{\frac{4}{\sqrt{3}}\phi_-}&=&\left(\frac{Q_2}{Q_1}\right)^{2/3}\frac{f_4}{f_1}=-\left(\frac{Q_2}{Q_1}\right)^{2/3}\frac{2}{64
    f_1f_2\Delta^4}~,\\
  b_+^4&=& 256 Q_1^2Q_2^2 f_2 f_5 \Delta^4~ = 4 Q_1^2Q_2^2 \left(\frac{f_2}{f_1}\right)~, \\
  b_-^4&=& 256 Q_1^2Q_2^2 f_1 f_4 \Delta^4~ = 4 Q_1^2Q_2^2 \left(\frac{f_1}{f_2}\right)~,
\end{eqnarray}
so that 
\begin{eqnarray}
  \label{eq:conc2}
  \phi_+ &=& \phi_-~,
\end{eqnarray}
and as expected
\begin{eqnarray}
  \label{eq:conc3}
  b_+^2b_-^2&=& 4 Q_1^2 Q_2^2=b_0^2~.
\end{eqnarray}

Now requiring asymptotic flatness and plugging in the asymptotic values of the scalars gives  (\ref{toda0}), (\ref{toda1})
\begin{eqnarray}
  \label{eq:f1f2f3}
  f_1+f_2+f_3 &=& \tfrac{1}{4}{(\bar Q_1)}^{-5/6}{(\bar Q_2)}^{-1/6}\\
    f_4+f_5+f_6 &=& \tfrac{1}{4}{(\bar Q_2)}^{-5/6}{(\bar Q_1)}^{-1/6}  \label{eq:f1f2f3b}
\end{eqnarray}
with $\bar Q_1 = e^{\sqrt 3 \phi_\infty}Q_1$ and $\bar Q_2 = e^{-\sqrt
  3 \phi_\infty}Q_2$. If we define
\begin{eqnarray}
  \label{eq:gdef}
  g_1&=&f_1+f_2~,\\
  g_2&=&f_1f_2~,
\end{eqnarray}
we can use (\ref{eq:const1}), (\ref{eq:const2}), (\ref{eq:const3}), (\ref{eq:const4}),
 (\ref{eq:f1f2f3}), (\ref{eq:f1f2f3b}) to find cubic equations for $g_1$ and $g_2$.
These are a little unwieldy, so 
for producing various plots, we found the following form of the solution useful~\cite{Gibbons:1985ac}:
\begin{eqnarray}
  \exp(4\phi/\sqrt{3}) & =&e^{4\phi_{\infty}/\sqrt{3}}  \frac{p_2}{p_1}~,\\
  a^{2} & = & \frac{(r-r_{+})(r-r_{-})}{\sqrt{p_1 p_2}}~,
\end{eqnarray}
where
\begin{equation}
  \label{pi}
  p_i=(r-r_{i+})(r-r_{i-})~,
\end{equation}
\begin{eqnarray}
  \label{ripm}
  r_{1\pm}&=&-\frac{1}{\sqrt\Sigma}\pm\bar Q_1\sqrt{\frac{2\Sigma}{\Sigma+\sqrt3M}}~,\\
    r_{2\pm}&=&\frac{1}{\sqrt\Sigma}\pm\bar Q_2\sqrt{\frac{2\Sigma}{\Sigma-\sqrt3M}}~,
\end{eqnarray}
and the scalar charge $\Sigma$ satisfies the cubic equation
\begin{equation}
  \label{eq:cubic}
  \frac{2}{3}\Sigma = \frac{\bar Q_1^2}{\Sigma+\sqrt3M}+\frac{\bar Q_2^2}{\Sigma-\sqrt3M}~.
\end{equation}
The relationship between the two forms of the solution shown above is
non-trivial because of the cubic equations that $\Sigma$ and $g_{1,2}$ satisfy.

\section{Perturbation}
\label{sec:pert}

We wish to construct a perturbation series about the non-extremal
constant scalar solution. As discussed in Section \ref{sec:mod}, when the scalars are
fixed to the minimum of $V_\mathrm{eff}$, we have a  Reissner--Nordstr\"om background which we can write as:
\begin{eqnarray}
  \label{eq:rn}
  a_0^2(r) = \frac{(r-r_+)(r-r_-)}{r^2}~, \quad b_0(r)=r~, \quad \phi=\phi_0~,
\end{eqnarray}
where $\p_\phi V_{\mathrm{eff}}|_{\phi_0}=0$.
From the constraint (\ref{constraint}) we obtain 
\begin{equation}
  \label{eq:bp0bm0}
b_{0+}b_{0-}=r_+r_-=V_{\mathrm{eff}}(\phi_0)=V_0~.  
\end{equation}

Suppose we now consider
a scalar near the attractor value, let
\begin{equation}
  \phi = \phi_0 + \epsilon \phi_1 + {\cal O}(\epsilon^2)~,
\end{equation}
Neglecting back-reaction, the equation for $\phi_1$ becomes
\begin{equation}
  \label{eq:pert2}
  ((r-r_+)(r-r_-)\phi'_1)'= \sigma^2\phi_1/(2 r^2)~,
\end{equation}
where $\sigma^2=\p_\phi^2V_\mathrm{eff}|_{\phi=\phi_0}$ is the coefficient of the first order expansion of the RHS of (\ref{3}).
 Now, substituting
\begin{equation}
  \label{eq:z}
  z= \left( \frac{r_+-r}{r}\right)\frac{r_-}{r_+-r_-}~,
\end{equation}
 into (\ref{eq:pert2}), which moves the poles  from $\{r_+,r_-,0\}$ to $\{0, 1,\infty\}$,  gives
\begin{equation}
  \label{eq:pert3}
  \partial_z(z(z-1)\partial_z\phi_1)=\lambda^2\phi_1~,
\end{equation}
where
\begin{equation}
  \label{eq:lam_deff}
  \lambda^2=\frac{\p_\phi^2V_\mathrm{eff}|_{\phi=\phi_0}}{2V_\mathrm{eff}|_{\phi=\phi_0}}~.
\end{equation}
For an effective potential of the form
$e^{\alpha_1\phi}Q_1^2+e^{\alpha_2\phi}Q_2^2$, it is not hard to see
that $\lambda^2=-\alpha_1\alpha_2/2$.  Now (\ref{eq:pert3}) has the
solution
\begin{equation}
  \label{eq:pert_sol}
  \phi_1(z)=c_1 P_\nu (2z-1)+c_2 Q_\nu(2z-1)~,
\end{equation}
where $P_\nu$ and $Q_\nu$ are Legendre functions of the first and
second kind respectively and
$\nu=\tfrac{1}{2}(\sqrt{1+4\lambda^2}-1)$.  If $\nu$ is
not an integer the perturbation diverges on both horizons since, for non-integer $\nu$,
$P_\nu(x)$ has a singularity at $x=-1$ which corresponds to the outer
horizon and $Q_\nu(x)$ has singularities at $x=\pm 1$ which
corresponds to the inner and outer horizon.

\section{Hamiltonian formulation for conservation laws}\label{appc}
Suppose we take  a static slice of the spacetime, bounded by a sphere at infinity and the event horizon and foliate it radially.
Then the dynamics governing the fields, evolving radially, on each hypersurface should follow the $tt$ component of the Einstein equation obtained by varying the action with respect to $g^{00}$.
The resulting Einstein's equation can be formulated as a conserved density: $G_{00} - 8 \pi G_D T_{00}= 0$. The resulting integral over the constraint manifold gives an effective ``Hamiltonian,'' which is null over the phase space of solutions of the system and whose variations with respect to the induced metric on the foliation yields the correct Hamiltonian equations on the foliation hypersurface.

For the static spherically symmetric solution we consider \cite{hep-th/0507096,hep-th/0512138}, the effective Hamiltonian density, as given in (2.49) of \cite{1108.0296},\footnote{Put $a = e^U$, $b =  e^A$, and $\psi  = A + U$. The extra $-1$ above arises from the positivity of the horizon curvature.} is
\begin{equation}
0= {\cal H} =  \left((a^2 b)'b' - a^2 b^2 (\phi')^2 - \frac{V_\mathrm{eff}(\phi)}{b^2}\right) - (a^2 (b^2)')' -1 ~.
\end{equation}
This relation applies to Region 1.
We may perform a variation in solution space as we move from one solution to another infinitesimally closer to it.
We make sure that these variations do not trigger any non-normalizable modes at either boundary; this is ensured by adding counterterms to cancel all such variations. The inner boundary, corresponding to the horizon at $r_+$, is defined by $a^2 =0$  while both $b^2$ and $(b^2)'$ diverge at the outer boundary, $r=\infty$.
This allows us to perform a variation of the background and matter fields subject to a null variation of the Hamiltonian density. As this condition is true at every point, one can integrate these null variations over a given time slice to obtain the total null variation:
\begin{equation}
 0 =  \left. \left\{ \delta(H_V) + \frac12 \left[ (b^2)' \delta a^2 + (a^2)' \delta b^2 + \frac{(a^2) (b^2)'}{2b^2} \delta b^2\right] - \delta  (a^2 (b^2)')- 2 a^2 b^2 (\phi')^i \delta \phi_i\right\} \right|^{\infty}_{r_+} ~. \label{c2}
\end{equation}
Here, the first term in the variation is the volume term which vanishes independently on shell. The surface terms, however, have a constrained variation as noted above.

Keeping in mind that the asymptotic form of the metric elements is
\be
a^2_{\infty} \approx 1 - \frac{2M}{r} + \frac{r_+ r_-}{r^2} ~, \qquad b^2_{\infty} = r^2 ~,
\ee and those of the scalar fields is
\be
\phi^i = \phi_0^i + \frac{\Sigma^i}{r} + \ldots ~,
\ee
we see that $\delta b^2$ nor $\delta (b^2)'$ cannot vary at infinity; the variation $\delta a^2$ at infinity is $-\frac{2\delta M}{r}$.
The variation $\delta a^2$, meanwhile, vanishes at the horizon.

Evaluating~\eqref{c2}, we find
\bea
0 &=& \left. -\left. \frac{(b^2)'}{2} \delta a^2 \right|_\infty - \left. \frac{(a^2)'}{2} \delta b^2 \right|_{r_+} - 2 a^2 b^2 (\phi')^i \delta\phi_i \right|_\infty \nn
&=& 2\delta M - 2 T_+ \delta S_+ + 2 \Sigma^i \delta \phi_i ~.
\eea
This is nothing but the first law for the outer horizon~\cite{hep-th/061114} at fixed charges.

Let us now repeat the exercise for Region 2.
From~\eqref{c2}, we deduce that
\be
0 = \left. \frac{(a^2)'}{2} \delta b^2 \right|^{r_+}_{r_-} = \frac12 (a^2)'_+ \delta b^2_+ - \frac12 (a^2)'_- \delta b^2_- ~. \label{qwe}
\ee
We know from~\eqref{6} that $(a^2)'_\pm = 4\pi T_\pm$.
This means that if the entropy is proportional to the horizon area ($S_\pm \propto b^2_\pm$), we may write
\be
T_+ \delta S_+ = T_- \delta S_- ~.
\ee
Using~\eqref{6} and~\eqref{8}, we see that
\be
\frac12 (a^2)'_\pm = \pm \frac{\Delta}{b^2_\pm} ~.
\ee
This allows us to recast~\eqref{qwe} as
\be
0 = \Delta \left( \frac{\delta b^2_+}{b^2_+} + \frac{\delta b^2_-}{b^2_-} \right) = \Delta \frac{\delta(b^2_+ b^2_-)}{b^2_+ b^2_-} ~.
\ee
The variation $\delta(b^2_+ b^2_-)$ should be zero in order to satisfy the Hamiltonian constraint at every point.
This implies that $b^2_+ b^2_-$ is constant.
To evaluate this constant, we go to a point in moduli space where the
asymptotic moduli are fixed at their attractor values so that we have
a Reisner-Nordstr\"om solution of the form \eqref{eq:rn}. 
From \eqref{eq:bp0bm0} we conclude that, $ b_+ b_- =  V(\phi_{0})$, 
which is independent of the asymptotic moduli. If the extremal solution is non-singular 
we can further conclude that
\be
b^2_+ b^2_- = b^4_\mathrm{ext} \qquad \Longleftrightarrow \qquad A_+ A_- = A_\mathrm{ext}^2 ~.
\ee

\bibliographystyle{utphys}
\bibliography{nonlocal}

\end{document}